\documentclass{aa}  
\usepackage{graphicx}
\usepackage{txfonts}
 \usepackage{aalongtable,lscape}
 \usepackage{colortbl}
 \usepackage{natbib}
\usepackage{url}
 \bibpunct{(}{)}{;}{a}{}{,}
\begin{document}
   \title{A robust morphological classification of high-redshift galaxies using support vector machines on seeing limited images}
   \subtitle{I. Method description
\thanks{Based on observations obtained at the Canada-France-Hawaii
    Telescope (CFHT) which is operated by the National Research
    Council of Canada, the Institut National des Sciences de l'Univers
    of the Centre National de la Recherche Scientifique of France, and
    the University of Hawaii.} 
}
   
\author{M. Huertas-Company
          \inst{1,4},
          D. Rouan \inst{1},
          L. Tasca \inst{3},
          G. Soucail \inst{2},
          O. Le F\`evre \inst{3}
          }

   \institute{LESIA-Paris Observatory, 
   5 Place Jules Janssen, 92195 Meudon, France\\
              \email{marc.huertas@obspm.fr}         
            \and
Laboratoire d'Astrophysique de Toulouse-Tarbes, CNRS-UMR 5572 and Universit\'e Paul Sabatier Toulouse III, 14 Avenue Belin, 31400 Toulouse,  France
            \and
           LAM-Marseille Observatory, Traverse du Siphon-Les trois Lucs BP8-13376 Marseille Cedex 12, France
           \and
           IAA-C/ Camino Bajo de Hu\'etor, 50 - 18008 Granada, Spain}

   \date{Received <date> / Accepted <date>}

 
  \abstract
   { Morphology is the most accessible tracer of galaxies physical
structure, but its interpretation in the framework of galaxy evolution
still remains a problem. Its dependence on wavelength turns indeed the
comparison between local and high redshift populations difficult.
Furthermore, the quality of the measured morphology being strongly
dependent on the image resolution, the comparison between different
surveys is also a problem. 
}
   {We present a new non-parametric method to quantify morphologies of
galaxies based on a particular family of learning machines called
support vector machines.  The method, that  can be seen as a
generalization of the classical CAS classification but with an
unlimited number of dimensions and non-linear boundaries between
decision regions, is fully automated and thus particularly well
adapted to large cosmological surveys. The source code is available for download at \url{http://www.lesia.obspm.fr/~huertas/galsvm.html}
}
   {To test the method, we use a seeing limited near-infrared ($K_s$
band, $2,16\mu m$) sample observed with WIRCam at CFHT at a median
redshift of $z\sim0.8$. The machine is trained with a simulated sample
built from a local visually classified sample from the SDSS chosen in
the high-redshift sample's rest-frame (i band, $0.77\mu m$ ) and
artificially redshifted to match the observing conditions. We use a
12-dimensional volume, including 5 morphological parameters, 
and other caracteristics of galaxies such as luminosity and redshift.
A fraction of the simulated sample is used to test the machine and assess its accuracy.  }
   {We show that a qualitative separation in two main morphological types (late type and early type) can be obtained with an error lower than $20\%$ up to the completeness limit of the sample ($KAB\sim 22$) which is more than 2 times better that what would be obtained with a classical C/A classification on the same sample and indeed comparable to space data. The method is optimized to solve a specific problem, offering an objective and automated estimate of errors that enables a straightforward comparison with other surveys. Selecting the training sample in the high-redshift sample rest-frame makes the results free from wavelength dependent effects and hence its interpretation in terms of evolution easier. }
   
   


\authorrunning{Huertas-Company et al}
\titlerunning{Morphology with SVM}
 \maketitle
%

\section{Introduction}

The process of galaxy formation and the way galaxies evolve is still one
of the key unresolved problems in modern astrophysics.  In the
currently accepted hierarchical picture of structure formation, galaxies
are thought to be embedded in massive dark halos that grow from density
fluctuations in the early universe \citep{Fa80} and initially contain
baryons in a hot gaseous phase. This gas subsequently cools, and some
fraction eventually condenses into stars \citep{Lil96,Mad98}. However,
many of the physical details remain uncertain, in particular the process
and history of mass assembly. One classical observational way to test
those models is to classify galaxies according to morphological criteria, i.e., the organization of its brightness as projected on the sky's plane and observed at a particular wavelength,
defined in the nearby Universe \citep{Hub36,deVauc48,San61}, and to follow this classification across
time \citep{Ab96,Sim02,Abr03}. 
Comparison of distant populations with
the ones found in the nearby Universe might help to clarify the formation history of the galaxy 
\citep{Co00, Ba96}. Progress in this field have
come from observing deeper and larger samples, but also from obtaining
higher spatial resolution at a given flux and at a given redshift.  In the
visible, progress has been simultaneous on those two fronts, thanks in particular to the
ultra-deep HDF fields observed with the Hubble Space Telescope. HST imaging has brought observational evidence that galaxy evolution
is differentiated with respect to morphological type and that a large
fraction of distant galaxies have peculiar morphologies that do not fit
into the elliptical-spiral Hubble sequence \citep{Bri98, Wo03, Il06}. 

However, a major obstacle is still the difficulty in quantifying morphology of high redshift objects with a few simple, reliable measurements. Indeed, with the increasing number of cosmological surveys available today, classical visual classifications become useless and automated methods must be employed. Globally there exist two main approaches: the first one, known as parametric, consists in modeling the distribution of light with an analytic model and fit it to the real galaxy. A commonly used parameter in this approach is the bulge-to-disk (B/D) light ratio that correlates with qualitative Hubble type classifications, and can be obtained by fitting a two-component profile \citep{Sim02, Peng02}. The main advantage of such a method is that the fitting output provides a quantitative morphology, i.e. a  complete set of parameters that describe the galaxy's shape (disk scale length, bulge effective radius...). Results are, unfortunately, often degenerated because of the high number of parameters to be adjusted \citep{Huer06}, even when the residuals are almost null, and the obtained accuracy strongly depends on the observing conditions (angular resolution, S/N...). Moreover this approach assumes that the galaxy is well described by a simple, symmetric profile, which is not true for irregular or well resolved objects.

The second approach is called non-parametric and basically consists in measuring a set of well-chosen parameters that correlate with the Hubble type. The main advantage of this method is that it does not assume a particular analytic model and can therefore be used to classify regular as well as irregular galaxies. The resulting morphology will be however more qualitative. \citet{Ab94, Ab96} first proposed this method by defining the concentration and asymmetry (C and A) parameters. They showed that plotting those values in a 2D plane, results in a quite good separation between the three main morphological types (early type, late type and irregulars). Subsequent authors modified then the original definitions to make C and A more robust to surface-brighntess selection, centering errors or redshift dependence \citep{Bri98,Wu99,Ber00,Con00} and introduced new parameters. In particular a third parameter the smoothness (S) was proposed by \cite{Con03} and gave its name to the CAS morphological classification system. More recently \cite{Abr03} and \cite{Lotz04} proposed two new parameters: the Gini coefficient that correlates with concentration and the M20 moment. Each of those parameters brings a different amount of information concerning the galaxy shape. There is no way, however, with classical approaches  to use more than 3 parameters simultaneously. \cite{Ber00} made a first attempt to do a multi-parameter analysis on a nearby sample using a 4 dimensional space including concentration and asymmetry as well as luminosity and color information. They found indeed correlations between those parameters and defined six 2D planes resulting from the combinations of those parameters. The classification was however done independently in each plane without considering all the information simultaneously. In the framework of the COSMOS consortia\citep{Sco05}, \cite{Scar06} have recently made a step forward by proposing a multi-parameter classification scheme (ZEST) based on the positions of galaxies in a three dimensional space resulting from a principal component analysis on a 5 dimensional space. The method uses almost all the information contained in the 5 parameters, but the final calibration is done in 3 dimensions.\\
Indeed, one key point in this kind of analysis is to correctly calibrate the volume, i.e. to correctly estimate the decision regions. One approach, is to use boundaries defined in the nearby universe from a visually classified sample and assume they will remain unchanged for a sample at high redshift, observed at a different wavelength and with an other instrument \citep{Ab96}. However, it is well known that the galaxy morphology depends on wavelength (K-correction) and on the observing conditions, that's why some corrections should be applied to take these effects into account \citep{Bri98}. Another approach consists in classifying visually a fraction of the sample and plot the boundaries according to the positions of galaxies in the space \citep{Men06, Scar06}. This of course takes into account the observing conditions of the sample but requires enough resolution and S/N to be able to decide visually the galaxy morphology. This is possible for space observations but becomes more difficult for ground-based observations, where the low resolution doesn't allow a reliable visual classification. In all these approaches, boundaries are forced to be linear (2D lines or hyper-planes) and are generally plotted manually which introduces a subjective element that turns more difficult a correct estimate of errors.




In this paper, we propose a generalization of the non-parametric classification that uses an unlimited number of dimensions and non-linear separators, enabling to use simultaneously all the information brought by the different morphological parameters. The approach uses a particular class of learning machines (called support vector machines) that finds the optimal decision regions in a volume using a training set. Here, we build this training set from a local sample that is transformed to reproduce the physical and instrumental properties of the science sample, allowing to use it even on seeing limited observations. The algorithm defines, in an automated way, the optimal decision regions using multi-dimensional hyper-surfaces as boundaries. It allows therefore a straightforward comparison between different science samples. The classification scheme that we propose is intended as a framework for future studies on large cosmological fields.

The paper proceeds as follows: generalities on pattern recognition and in particular on support vector machines (SVM) are described in the next section. In Section~\ref{sec:nearby} we make sure that SVM works properly when applied to a nearby sample. In Section~\ref{sec:high_z}, we describe the general steps of the proposed method to classify high-redshift objects. We show, in particular, how the training set is built to reproduce the real sample properties (\ref{sec:method}), we define the parameters measured for the morphological classification (\ref{sec:morpho_par}) and we finally describe several tests performed to probe the accuracy of the method (\ref{sec:tests}). 

We use the following cosmological parameters throughout the paper: $H_0 = 70\,\mathrm{km\,s^{-1}\,Mpc^{-1}}$ and
$(\Omega_\mathrm{M}, \Omega_\Lambda, \Omega_\mathrm{k})=(0.3,0.7,0.0)$.

\section{Generalities on pattern recognition}

Suppose a set of observations of a given phenomenon, in which each observation consists of a vector ${\bf x_i} \in \mathbb R^{n}, i=1,...,l$ and of an associated "truth" $y_i$. For instance, in a classical concentration and asymmetry classification plane, $\bf{x_i}$ would be a 2D vector whose components are the concentration and the asymmetry, and $y_i$ would be 0 if the galaxy is irregular, 1 if it is disk dominated and 2 if it is bulge dominated. We then call learning machine, a machine whose task is to learn the mapping ${\bf x_i}\mapsto y_i$ defined by a set of possible mappings ${\bf x}\mapsto f({\bf x},\alpha)$. A particular choice of $\alpha$ generates what is called a "trained machine".

\subsection{Support vector machines}
Support vector machines are a particular family of learning machines, first introduced by \cite{Vap95} as an alternative to neural networks and that have been successfully employed to solve clustering problems, specially in biological applications. 

\begin{figure}
 \centering 
  \resizebox{\hsize}{!}{\includegraphics{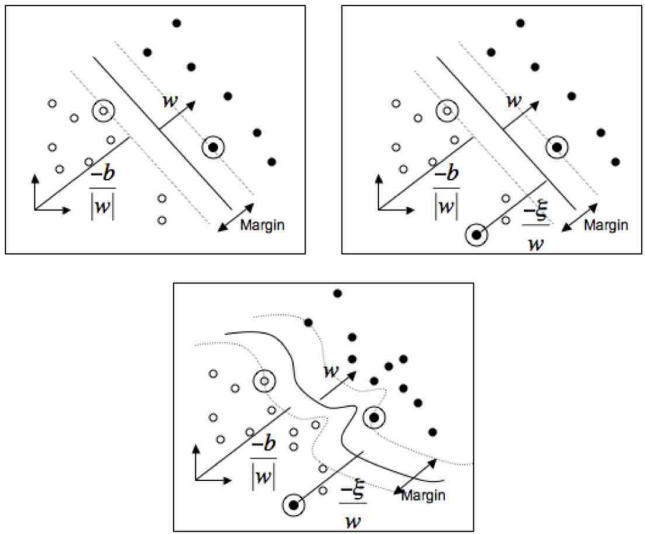}}
 \caption{2D illustration of the three cases of SVM classification. Top left: linearly separable data with linear boundaries. Top right: Non-linearly non-separable data with linear boundaries. Bottom: Non-linealry non-separable data with non-lindear borders.} 
 \label{fig:svm_scenario} 
 \end{figure}

In order to simplify the description of the most important points concerning SVM we will focus on a 2 class classification problem: $\{ {\bf x_i}, y_i\}, i=1,...,l$ $y_i \in \{-1,1\}, {\bf x_i} \in \mathbb R^d$ without loss of generalization.
The basic idea is to find an hyperplane that separates the positive from the negative examples. If this plane exists, the points $\bf x$ that lie on the hyperplane satisfy ${\bf w.x} +b=0$, with $\bf w$ normal to the hyperplane, and $|b|/\|{\bf w}\|$ the perpendicular distance from the hyperplane to the origin. $d_+(d_-)$ will then be the shortest distance from the separating hyperplane to the closest positive (negative) example. The "margin" is defined to be: $d_++d_-$. The algorithm will then simply look for the separating hyperplane with largest margin. 
This can be formulated as follows:
\begin{enumerate}
\item ${\bf x_i.w} +b \geq +1$ for $y_i=+1$
\item ${\bf x_i.w} +b \leq-1$ for $y_i=-1$
\end{enumerate}
The training points for which the equalities hold and whose removal would change the solution are called support vectors (Figure~\ref{fig:svm_scenario}).

It can be (and it is the most common case) that it does not exist a linear hyperplane that perfectly separates the two data sets. In this case we can relax constraints by introducing a positive stack variable $\xi_i,$ $i=1,...,l$ and the equations become then:
\begin{enumerate}
\item ${\bf x_i.w} +b\geq +1-\xi_i$ for $y_i=+1$
\item ${\bf x_i.w} +b \leq-1+\xi_i$ for $y_i=-1$
\end{enumerate}
The global effect is to change the objective function to be minimized to $\|w\|^2/2+C(\sum \xi_i)$, where C is a parameter to be chosen by the user, a larger C corresponding to assigning a higher penalty to errors.

Another feature that can be added to solve more complex problems are non linear decision functions. To do so, we map the data to some other (possibly infinite dimensional) Euclidian space $H$: $\Phi: \mathbb R^d \mapsto H$ where the data can be linearly separable by some hyperplane. Since the only way in which the data appear in the training problem is in the form of dot products ${\bf x_i.x_j}$ then the training algorithm would only depend on the data through dot products in $H$, i.e. on functions of the form $\Phi (x_i).\Phi (x_j)$. If there is a "kernel function" K such that $K(x_i,x_j)=\Phi (x_i).\Phi (x_j)$ we would never need to explicitly even know what $\Phi$ is. Examples of kernels are: $K(x,y)=(x.y+1)^p (Polynomial)$, $K(x,y)=e^{-g\|x-y\|^2}(Gaussian RBF)$.

In summary, SVM are a particular family of learning machines that:

\begin{itemize}

\item for linearly separable data, simply look for the optimal separating hyperplane between distributions by maximizing the margin.

\item for non separable data a "tolerance" parameter C must be added which controls the tolerance to errors. 

\item for non linear non separable data a kernel function is built that maps the space into a higher dimensional space where the data are linearly separable. Then the Kernel parameters must be adjusted too. 

\end{itemize}

\subsection{Application to galaxies}
\cite{Ab94} proposed the idea of measuring some well-chosen parameters on a galaxy image that can be easily correlated with its morphology. In their paper they introduced the concentration, which basically measures the fraction of light contained in an inner isophote, and the asymmetry, which measures the degree of symmetry of the galaxy. They showed, that plotting those values in a 2D plane results in a quite good separation between the three main morphological populations: early-type, late-type and irregulars. They consequently plotted linear separators to define the regions and classified a set of galaxies with unknown morphology according to their positions in the so-called C/A plane. In other words, they tried to maximize the margins between 3 populations in a 2 dimensional space using linear separators. 
The same task can be done in a 3 dimensional space (CAS, \citealp{Con03}) but it becomes simply impossible with more than 3 dimensions. In this sense SVM offer a straightforward generalization of this method since they can separate samples with an unlimited number of dimensions and use non-linear boundaries.

Previous works \citep{Ab96,Bri98} have shown that morphological classification is far from being a linearly separable problem, since the contamination in the C/A plane is quite high. We have chosen therefore to use the most general SVM, i.e. a non linear machine with an RBF kernel. A machine is thus parameterized with two parameters: the tolerance (T) and the kernel exponential factor (g). Each possible combination of those two parameters generates a family of functions $f_{T,g}(\alpha,x_i)$. T and g must be fixed before performing the training and $\alpha$ is the result of the training procedure. There exist several techniques for finding the best T and g values for a  given problem; here we will use a cross-validation method described in \cite{CC01a}. It simply consists in performing a systematic search over a grid of possible values and selecting the pair that gives the best results.

Our goal is therefore to train a support vector based machine to estimate the morphology of a high redshift sample. We use throughout the paper the free available package \emph{libSVM} \citep{CC01a}. The procedure is basically the same as in a classical C/A classification but using a trained SVM to plot the optimal boundaries. As we show below, this allows to use more than two morphological parameters simultaneously and also to measure errors in an automated and objective way, which is capital to compare different classifications.

\section{Test on a well-resolved nearby sample}

\label{sec:nearby}
\subsection{Classical C/A classification versus 2-D SVM}

In order to verify that SVM work properly when applied to morphological classification of galaxies, we start with a simple test, i.e. classifying a local sample from the Sloan Digital Sky Survey (SDSS) in the i band that has been visually classified \citep{Tasc06}. Galaxies are nearby and consequently well-resolved and with a high S/N. Classical C/A classifications have been proved to give good results in such cases (e.g \citealp{Ab96, Men06}), therefore the idea is to verify that we get at least the same results using SVM and that no extra-biases are introduced. 

\begin{figure} 
 \centering 
  \resizebox{\hsize}{!}{\includegraphics{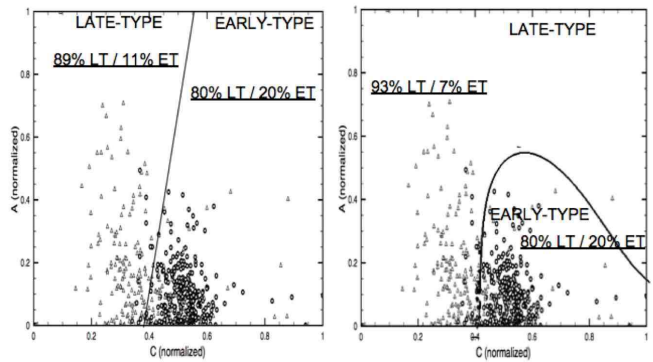}}
 \caption{C/A classical classification (left) versus C/A SVM classification (right) of 500 nearby objects. Triangles are galaxies visually classified as late-type and circles are galaxies visually classified as early-type. Numbers show the probability that the predicted morphological type is the same as the visual one.} 
 \label{fig:ca_class_svm_sdss} 
 \end{figure}

We thus measure concentration and asymmetry parameters (see ~\ref{sec:morpho_par} for details) and, on the one hand we try to plot the best linear boundary by eye as usually done to separate galaxies in two classes (late type and early type); on the other hand we train a SVM and finally, we compare the outputs. Figure~\ref{fig:ca_class_svm_sdss} shows the two resulting boundaries. The shape of the boundaries are quite different since SVM does not produce a linear boundary but when looking at the global accuracy we see that both methods are fully consistent. Indeed, the completeness (the fraction of visual classified galaxies that are correctly recovered) and the contaminations (the fraction of visual classified galaxies that are misclassified) are practically the same for the two methods (see Table~\ref{tbl:accu_comp_sdss}).

To confirm this consistency and to verify that no extra biases are introduced, we also made a one-to-one comparison of all the galaxies classified with the two methods. We obtain that $98\%$ ($94\%$) of the early-type (late-type) galaxies classified with the classical C/A method are also classified as early-type (late-type) using the trained SVM. 

We conclude that for a high S/N well-resolved sample, such as the SDSS sample, the use of SVM to plot the boundaries is equivalent to use classical procedures. The major advantage, however, is that the boundary is plotted automatically to minimize the errors.

\subsection{Classical C/A classification versus 4-D SVM}

Since one of the main advantages of using SVM is that it can work with an unlimited number of parameters, we investigate the effect of adding dimensions to the SVM classification. We thus classify the same sample as above but with four morphological parameters instead of two: Concentration, Asymmetry, Smoothness and Gini (see ~\ref{sec:morpho_par} for details on how they are calculated) and compare the outputs. Results are shown in table~\ref{tbl:accu_comp_sdss}. We see that there is no significant gain for this particular case. This suggests that, as proven in previous works (e.g \citealp{Ab96, Men06}), when dealing with a well-resolved and high S/N sample, concentration and asymmetry are enough to obtain an accurate morphological classification.

\begin{table*}[h!,b!]
\begin{center}
\begin{tabular}{c|cc|cc|cc|}
\hline\hline\noalign{\smallskip}

 & \multicolumn{2}{c|}{Classical C/A} & 
            \multicolumn{2}{c|}{SVM C/A} &
            \multicolumn{2}{c|}{SVM 4-D} \\ 
      
 & Early-Type & Late-Type & Early-Type & Late-Type & Early-Type & Late-Type  \\
\noalign{\smallskip}\hline\noalign{\smallskip}
Visual Early-Type &   0.80 (254) &0.09 (17) & 0.79 (256) & 0.08 (15) & 0.79 (251) & 0.10 (20) \\
Visual Late-Type &   0.20 (65) &0.91 (172)& 0.21 (72) & 0.92 (166)& 0.21 (67) & 0.90 (171) \\
\noalign{\smallskip}\hline
\end{tabular}
\end{center}
\caption{Comparison of the accuracy of three classifications of the SDSS sample: Classical C/A, SVM C/A and 4-D SVM. The table shows for each method the relations between the visual and the predicted morphological classes.  The number of objects are enclosed in parentheses. (see text for details) }
\label{tbl:accu_comp_sdss}
\end{table*}

\section{Going to higher redshift...}
\label{sec:high_z}

When observing objects at higher redshift with a ground-based telescope the S/N decreases, galaxies become poorly resolved and consequently more symmetric and less concentrated (e.g. \citealp{Con00}). The separation in the C/A plane turns out to be less clear. That's why space data such as HST imaging are widely used for those purposes and classifications based on colors are usually adopted for ground-based data (e.g. \citealp{Zucca06}). It is known however (e.g. \citealp{arn07}) that a classification based only on colors is highly contaminated by the presence, for instance, of an important population of  "blue" early-type galaxies, specially at high redshift where the red sequence is building up. That is one of the reasons why classifications based on morphological criteria are preferred. 
Indeed, with the increasing amount of data coming from ground-based surveys becoming available today it would be interesting to know if it is possible to obtain at least a rough morphological classification from these observations. 
In the following sections we therefore investigate wether the possibilities of using a large number of parameters and non-linear boundaries offered by support vector machines can help to increase the accuracy of "pure" morphological classifications on high-redshift ground-based data.

\subsection{Description of the employed method}
\label{sec:method}

\begin{figure} 
 \centering 
  \resizebox{\hsize}{!}{\includegraphics{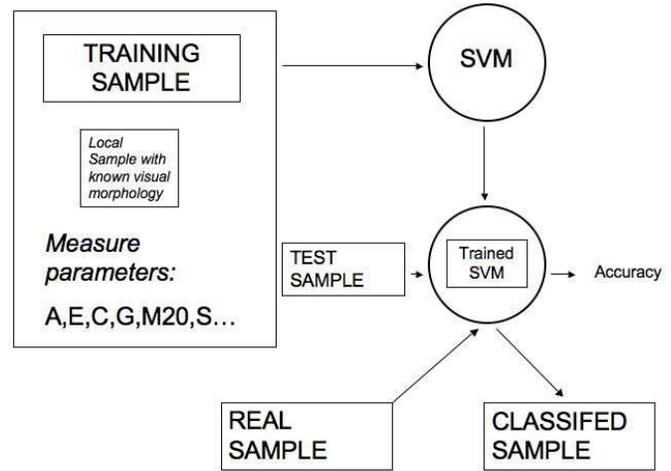}}
 \caption{Steps for morphological classification (see text for details).} 
 \label{fig:mag_z_dist} 
 \end{figure}

The proposed procedure can be summarized in 4 main steps (Figure~\ref{fig:mag_z_dist}):

\begin{enumerate}

\item Build a training set: for that purpose, we select a nearby visually classified sample at a wavelength corresponding to the rest-frame of the high redshift sample to be analyzed. We then move the sample to the proper redshift and image quality and drop it in the high z background. This is fully described in Section~\ref{sec:training}.

\item Measure a set of morphological parameters on the sample.

\item Train a support vector based learning machine with a fraction of the simulated sample and use the other fraction to test and estimate errors.

\item Classify real data with the trained machine and correct for possible systematic errors detected in the testing step.

\end{enumerate}

In the following sections, we describe each of the steps enumerated above.

\subsection{The training set}
\label{sec:training}

The most important step in obtaining the morphology with a non-parametric method is to correctly calibrate the volume filled by the data in the multi-dimentional space. This is a critical step since it will determine the decision regions that will be used to perform the classification. Indeed, galaxy morphology depends on the physical properties of the galaxy (luminosity, redshift, wavelength) and on the observing conditions (background level, resolution). A suitable calibration set should consequently reproduce closely all the properties of the sample to be analyzed. One classical approach consists in visually classifying a fraction of the sample and use it as a training set to optimize boundaries \citep{Men99,Men06}. However this is not possible for seeing limited data where the resolution is too low to enable a reliable visual classification. Here, we then decide to simulate the high redshift sample from a visually classified local catalog, selected in the rest-frame color of the high redshift sample. This has three main advantages: first, it's free from K-correction effects, second it does not introduce any modeling effect, since the used galaxies are real and finally, the training set is built to reproduce the observing and physical properties of the sample to be analyzed, but it is classified locally, so it does not need to have a specially high resolution.

\subsubsection{Real sample}

\begin{figure} 
 \centering 
  \resizebox{\hsize}{!}{\includegraphics{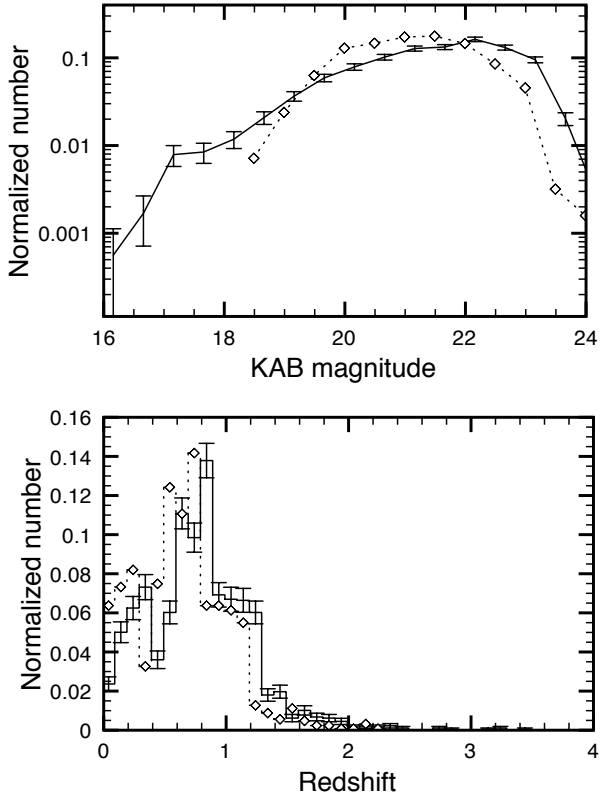}}
 \caption{Magnitude and redshift distributions of the real and simulated sample. Solid line: real sample. Dotted line: simulated sample. Error bars show poissonian errors for the real sample. See text for explanations concerning the differences between the simulated and real distributions.} 
 \label{fig:mag_z_wircam} 
 \end{figure}

In order to test the method, we work on a sample of galaxies
observed with WIRCam at CFHT in the near infrared $K_s$ band. The
field is part of the Canada-France Hawaii Telescope Legacy Survey
(CFHTLS) Deep survey and its near infrared follow-up and it is centered
on the COSMOS area \citep{Sco05}. We use a cutout of
$10^{'}\times10^{'}$ to perform all the tests. The sample is complete
up to $K(AB)=22$ and the median photometric redshift is $\sim 0.8$
(Fig.~\ref{fig:mag_z_wircam}). Images are reduced wit the Terapix
pipeline\footnote{http://terapix.iap.fr} and have a pixel scale of
$0.15^{"}$ with a mean FWHM of $0.7^{"}$. These data are particularly
interesting because K-band data have the advantage of probing old
stellar populations in the rest-frame, enabling a determination of
galaxy morphological types unaffected by recent star formation. Moreover, no space data in this wavelength range are available today.
Photometric redshifts come from the publicly available catalogue from
\cite{Ilb06b} computed with the
\emph{LePhare}\footnote{http://cencos.oamp.fr/cencos/CFHTLS/} code on
the CFHTLS Deep survey Terapix release T0003 and its multi-color photometric
catalogs. 

\subsubsection{Building the sample}

We used therefore a local catalog of $1472$ objects from the Sloan Digital Sky Survey in the i band, which roughly corresponds to the rest-frame of the K-band at $z\sim1$ and that has been visually classified \citep{Tasc06}. 

We first generate a random pair of (magnitude, redshift) values with a probability distribution that matches the real magnitude and redshift distribution of the sample to be simulated (see Figure~\ref{fig:mag_z_wircam}).

Then, for every galaxy stamp, we proceed in four steps:

\begin{enumerate}
\item First, we remove all the foreground stars and all other sources that do not belong to the galaxy itself. We use for that purpose the SExtractor segmentation map \citep{Ber96} and replace all the surrounding sources with a random noise with same statistics (mean value and variance) than the real background noise. The foreground stars that fall within the galaxy, are replaced with the mean value in the galaxy area.

\item Second, we degrade the resolution to reach the one at high redshift: we measure the FWHM at high redshift ($f_{hz}$), convert it to Kpc using a standard $\Lambda$CDM cosmology and deduce the resolution the local galaxy must have ($f_{lz}$). Then the image is convolved with a 2D gaussian function of $FWHM=\sqrt{(f_{lz}^2-f_i^2)}$, where $f_i$ is the local galaxy's initial resolution.  

\item Third, the image is binned to reach the expected angular size at high redshift with the $0.15^{"}$ pixel scale. In this step, the image is also scaled to its new magnitude. In the scaling procedure we force the final mean background level of the simulated stamp to be at least 3 times lower than the real background. This is to avoid that the local noise dominates over the high-redshift noise when dropping the galaxy in a real background. This implies that too bright objects (typically $K_s<17$ ) cannot be simulated since the necessary scaling factor is too small and explains the difference between the real and the simulated magnitude distribution in figure~\ref{fig:mag_z_wircam}. The difference in the faint end is due to the fact that some simulated objects are not detected by SExtractor.

\item Finally, we drop the galaxy in a real background image.
\end{enumerate}

\begin{figure} 
 \centering 
  \resizebox{\hsize}{!}{\includegraphics{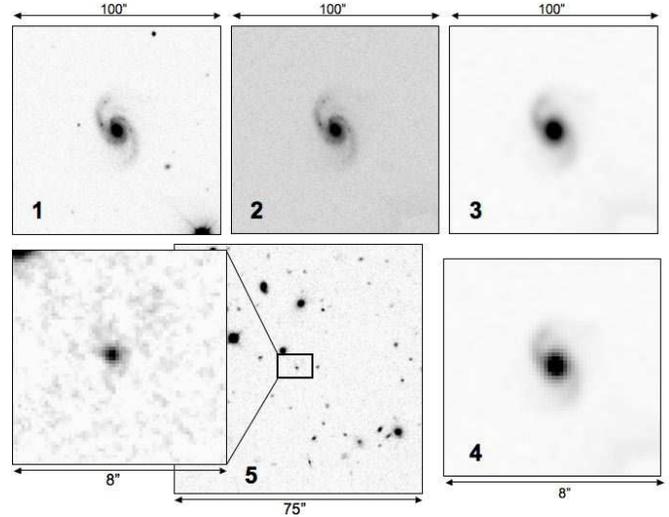}}
 \caption{Example of simulation for a galaxy. 1: SDSS i band image; 2: image after subtraction of foreground stars; 3: image after convolution; 4: image after binning; 5: final simulated field with real and simulated galaxies.} 
 \label{fig:dropping} 
 \end{figure}

Figure~\ref{fig:dropping} illustrates the entire procedure for a spiral galaxy.

In summary, we simulate a high redshift sample from a local sample selected in the high redshift sample's rest-frame to avoid K-correction effects. The sample reproduces the observing conditions (background level, noise, resolution) and physical properties (redshift and magnitude distribution) of the sample to be analyzed.

\subsection{Measuring morphological parameters}
\label{sec:morpho_par}

Once the simulated galaxies are dropped in a real background, we measure the following 5 morphological parameters:

 \begin{itemize}
 
 \item Concentration: basically, it measures the ratio of light within a circular or elliptical inner aperture to the light within a circular or elliptical outer aperture. Generally, it's defined in slightly different way by different authors. Here we adopt the \cite{Ber00} definition as for the ratio of the circular radii containing $20\%$ and $80\%$ of the "total flux": 
 \begin{equation}
 C=5log(r_{80}/r_{20}) 
 \end{equation}
 We use Conselice's (2003) \nocite{Con03} definition of the total flux as the flux contained within $1.5r_p$ (Petrosian radius). For the concentration measurement, the galaxy's center is that determined by the asymmetry minimization (see below).
 
 \item Asymmetry: it quantifies the degree to which the light of a galaxy is rotationally symmetric. It is measured by subtracting the galaxy image rotated by $180^{o}$ from the original image: 
  \begin{equation}
 A = \frac{1}{2}\left(\frac{\sum |I(i,j)-I_{180}(i,j)|}{\sum I(i,j)} - \frac{\sum |B(i,j)-B_{180}(i,j)|}{\sum I(i,j)}\right) ,
   \end{equation}
 where I is the galaxy image and $I_{180}$ is the galaxy image rotated by $180^{¡}$ about the galaxy's central pixel and B is the average asymmetry of the background. The central pixel is determined by minimizing A.
 
 \item Smoothness: developed by \cite{Con00}, it quantifies the degree of small-scale structure. The galaxy image is smoothed by a boxcar of given width and then subtracted from the original image: 
  \begin{equation}
 S = \frac{1}{2}\left(\frac{\sum |I(i,j)-I_{S}(i,j)|}{\sum I(i,j)} - \frac{\sum |B(i,j)-B_{S}(i,j)|}{\sum I(i,j)}\right),
  \end{equation}
  where $I_S$ is the galaxy's smoothed by a boxcar of width $0.25r_p$. 
 
 \item Moment of Light: introduced by \cite{Lotz04}, the total
second-order moment $M_{tot}$ is the flux in each pixel $f_i$
multiplied by the squared distance to the center of the galaxy, summed
all over the galaxy pixels assigned by the SExtractor segmentation
map: $M_{tot}=\sum{f_i[(x_i-x_c)^2+(y_i-y_c)^2]}$, where $x_c$ and
$y_c$ is the galaxy's center. The second-order moment of the brightest
regions of the galaxy traces the spatial distribution of any bright
nuclei, bars, spiral arms and off-center star clusters. We define
$M_{20}$ as the normalized second-order moment of the $20\%$ brightest
pixels of the galaxy.
 
 \item Gini Coefficient: it is a statistic based on the Lorentz curve, i.e. the rank-ordered cumulative distribution function of a population's wealth or in this case a galaxy's pixel values \citep{Abr03}. For the majority of local galaxies, the Gini coefficient is correlated with the concentration index and increases with the fraction of light in a central component. However, unlike C, G is independent of the large-scale spatial distribution of galaxy's light. Therefore, G differs from C in that it can distinguish between galaxies with shallow light profiles (which have both low C and G) and galaxies where much of the flux is located in a few pixels but not at the center (which have low C but high G).

\end{itemize}

Each of the above parameters, measure different properties of a galaxy and give therefore a different amount of information concerning the galaxy's morphological type. For instance, \cite{Lotz04} used the $M_{20}$/Gini plane to identify merger candidates whereas the C/A plane is classicaly used to separate late from early type galaxies. A multi-dimensional analysis allows consequently to use simultaneously all the information brought by each of the morphological parameters to increase the accuracy of the classification. Moreover, previous works have shown that  the measured parameters might also depend on the size, the luminosity or the redshift of the galaxy \citep{Bri98, Ber00}. Therefore, including non-morphologcial parameters should help the machine to take into account systematic trends in the morphological parameters due to luminosity or size variations. We thus measure 7 more parameters that we distribute in 4 classes: shape, size, luminosity and distance, according to the kind of information they measure:

\begin{itemize}

\item Shape: we include 2 shape parameters: the  galaxy ellipticity as measured by SExtractor, i.e. the ratio of the minor and major axis of the isophotal ellipses describing the galaxy, and the \textsc{CLASS\_STAR} parameter also from SExtractor. This parameter is intended to separate galaxies from stars and results from a neural network classification. Since it spans a continuum range between 0 and 1, it can be interpreted as a measure of the galaxy's compactness.

\item Size: the size parameters include the isophotal galaxy area and the petrosian radius. 

\item Luminosity: we use the apparent magnitude of the galaxy and the mean surface brightness.

\item Distance: we adopt the photometric redshift as a measure of the distance.

\end{itemize}

\subsection{Training and testing}

\label{sec:tests}
We perform several tests to probe the accuracy of the proposed method. For all the tests we adopt the same procedure: we use a fraction of the simulated catalogue (typically 500 galaxies) to train the machine and the remaining ~1000 objects to test it by looking at the fraction of galaxies that are correctly classified. We limit the analysis to only 2 broad morphological classes (late type and early type). The main reason for this choice is that there are too few irregular galaxies in the employed local sample to define a class. There is however no loss of generalization and the same analysis can be performed with an unlimited number of classes, provided of course that they are correlated with measured parameters. 

\subsubsection{Classical C/A classification versus 2-D SVM}

The first point we try to answer is how good would be the classification of this sample using a classical linear C/A classification. We thus get the brightest objects of the sample (with known visual morphology) ($K_s < 20$) and try to plot a linear boundary between the two distributions.
As expected, the distributions are now poorly separated and plotting a linear boundary becomes extremely difficult.
This is confirmed when trying to classify the whole sample (Table~\ref{tbl:accu_comp}): the completeness and contaminations are basically the same that we would have obtained with a random choice. We conclude that concentration and asymmetry alone cannot be used on this sample to obtain a reliable morphological classification. 

In a second step, we classify this sample with a SVM machine with the same two parameters. Results are shown in table~\ref{tbl:accu_comp}. We observe, a slight gain due to the fact that SVM can adapt boundaries in a non-linear way, but the accuracy is still comparable with a random choice.

\subsubsection{n-D versus 2-D SVM}
\label{sec:nd_2d}
\paragraph{Global effect}

We trained then 2 machines: the first one, with only 2 parameters (C and A), which should globally give the same results as a classical C/A classification as shown in section~\ref{sec:nearby} and the second one with 12 parameters described above. We then tested both machines by looking at the fraction of galaxies that are correctly classified. Results for the whole sample are summarized in table~\ref{tbl:accu_comp}. We observe that including more than two parameters in the classification results in a significant gain for this sample where C/A cannot do much better than a random choice. Indeed we almost recover the same accuracy that was obtained on the nearby sample (Table ~\ref{tbl:accu_comp_sdss}).

\begin{table*}[h!,b!]
\begin{center}
\begin{tabular}{c|cc|cc|cc|}
\hline\hline\noalign{\smallskip}

 & \multicolumn{2}{c|}{Classical C/A} & 
            \multicolumn{2}{c|}{SVM C/A} &
            \multicolumn{2}{c|}{SVM 12-D} \\ 
           
 & Early-Type & Late-Type & Early-Type & Late-Type & Early-Type & Late-Type  \\
\noalign{\smallskip}\hline\noalign{\smallskip}
Visual Early-Type &   0.59 (96) &0.51 (321) & 0.57 (304) & 0.45 (113) &0.75 (365) & 0.18 (52) \\
Visual Late-Type &   0.41 (65) &0.49 (309)& 0.43 (236) & 0.55 (138)&0.25 (149) & 0.82 (225) \\
\noalign{\smallskip}\hline
\end{tabular}
\end{center}
\caption{Comparison of the accuracy of three classifications of the WIRCam sample: Classical C/A, SVM C/A and 12-D SVM. The table shows for each method the relations between the visual and the predicted morphological classes. The number of objects are enclosed in parentheses. (see text for details) }
\label{tbl:accu_comp}
\end{table*}

\paragraph{Robustness}

We now try to establish the robustness of this effect.
\begin{figure*}[h!t!]
\begin{center}
$\begin{array}{c@{\hspace{1in}}c}
	\includegraphics[width=0.38\textwidth,height=0.3\textwidth]{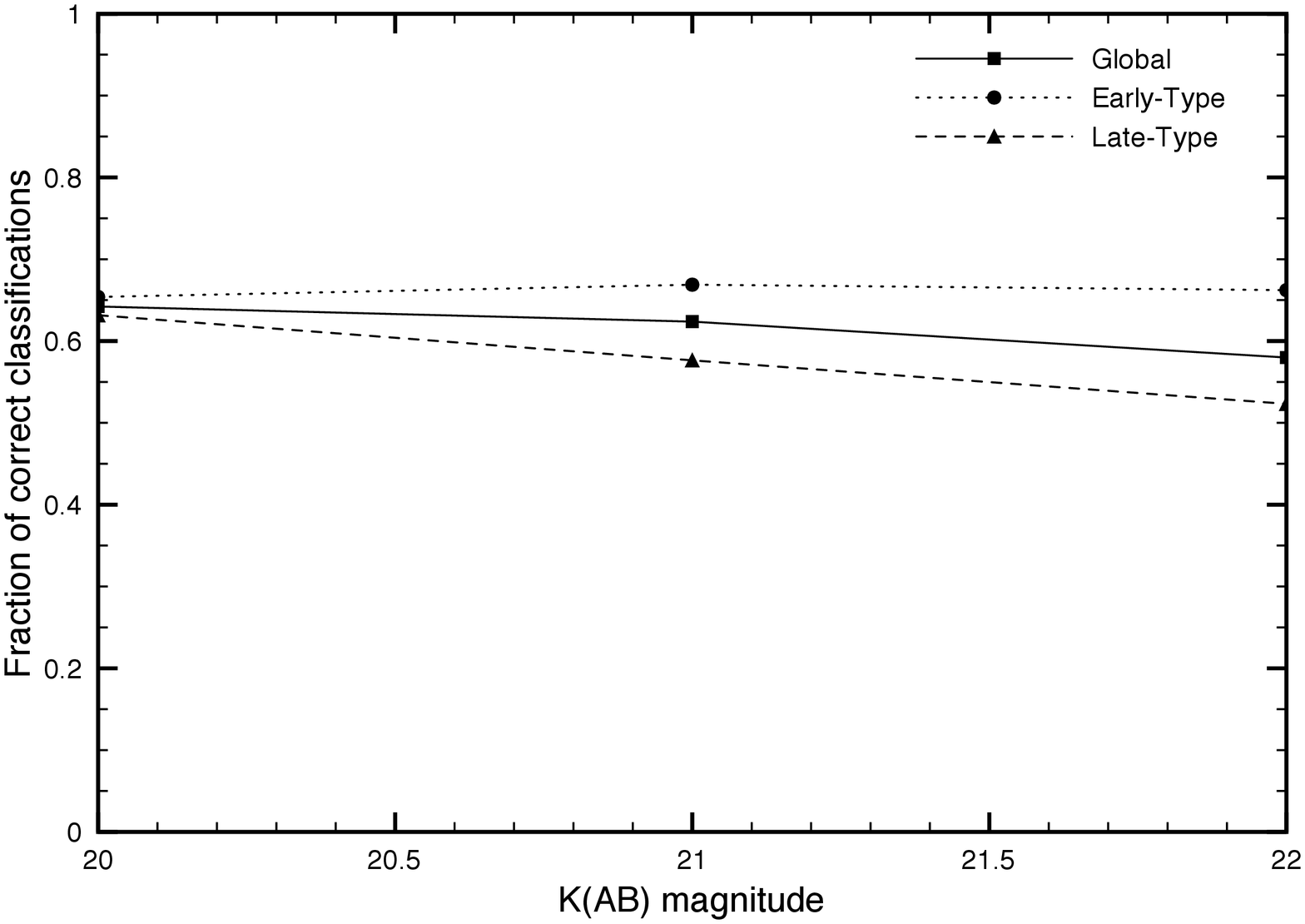} &
	\includegraphics[width=0.38\textwidth,height=0.3\textwidth]{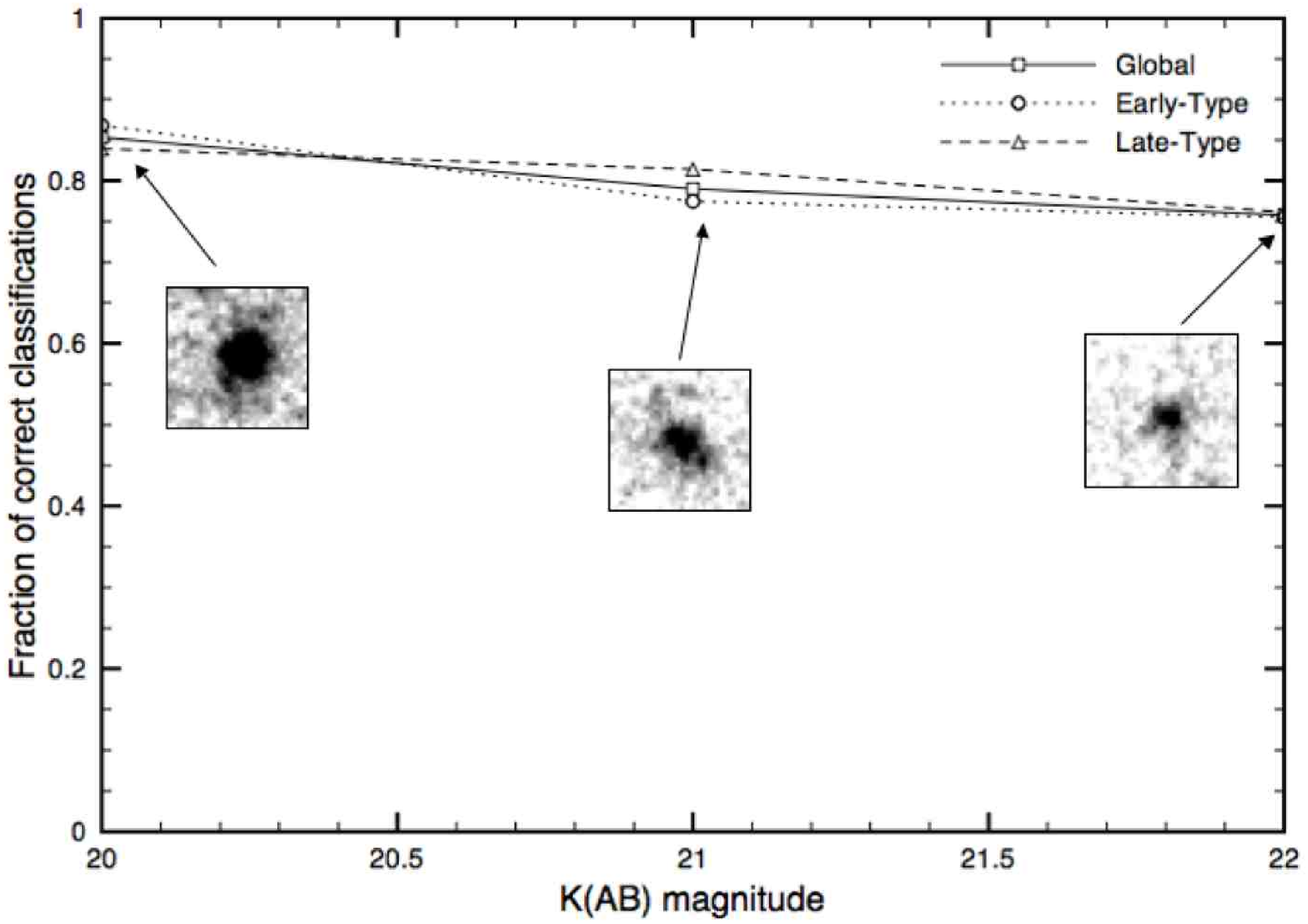} \\
\mbox{\bf (a)} & \mbox{\bf (b)}\\
	\includegraphics[width=0.38\textwidth,height=0.3\textwidth]{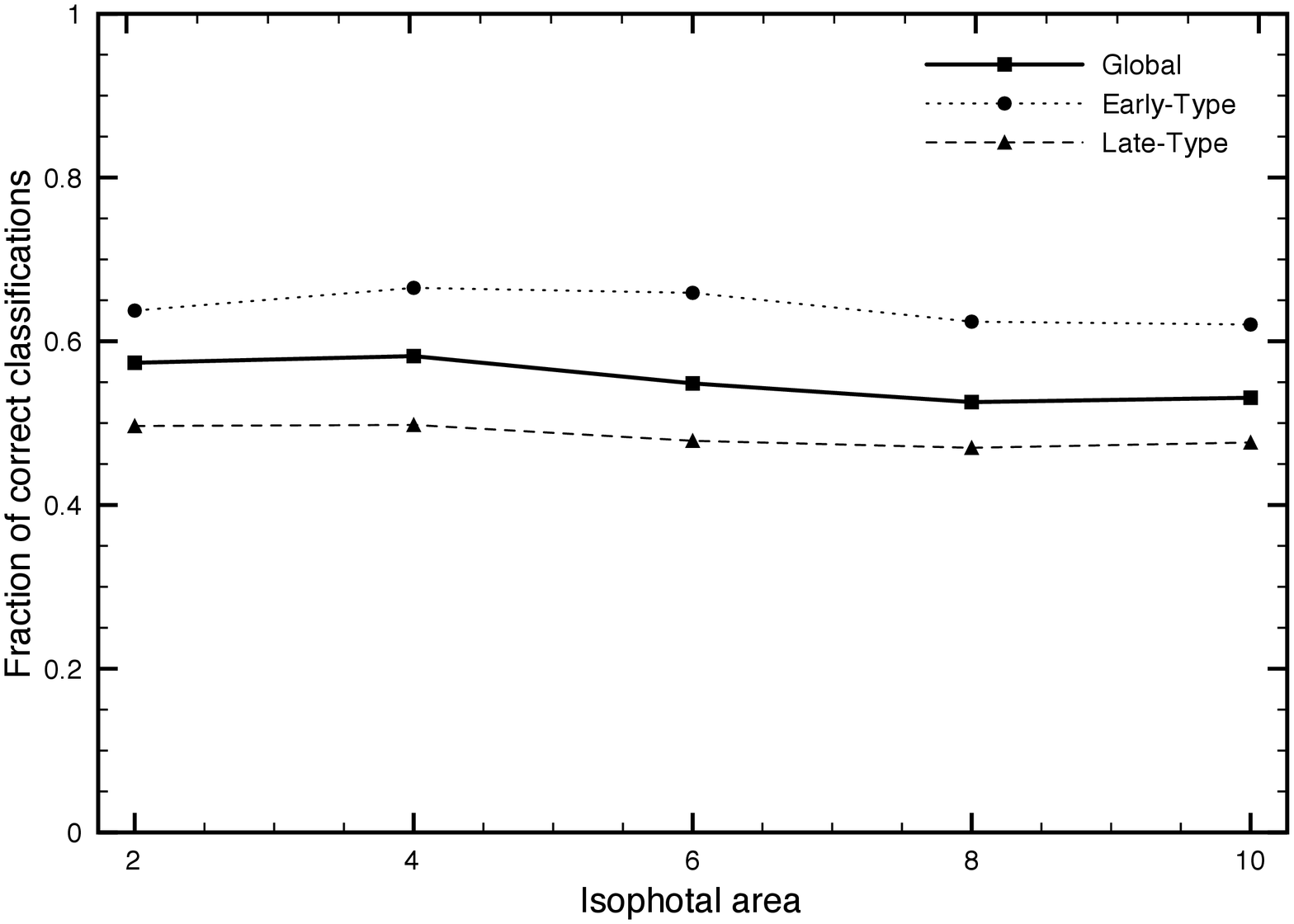} &
	\includegraphics[width=0.38\textwidth,height=0.3\textwidth]{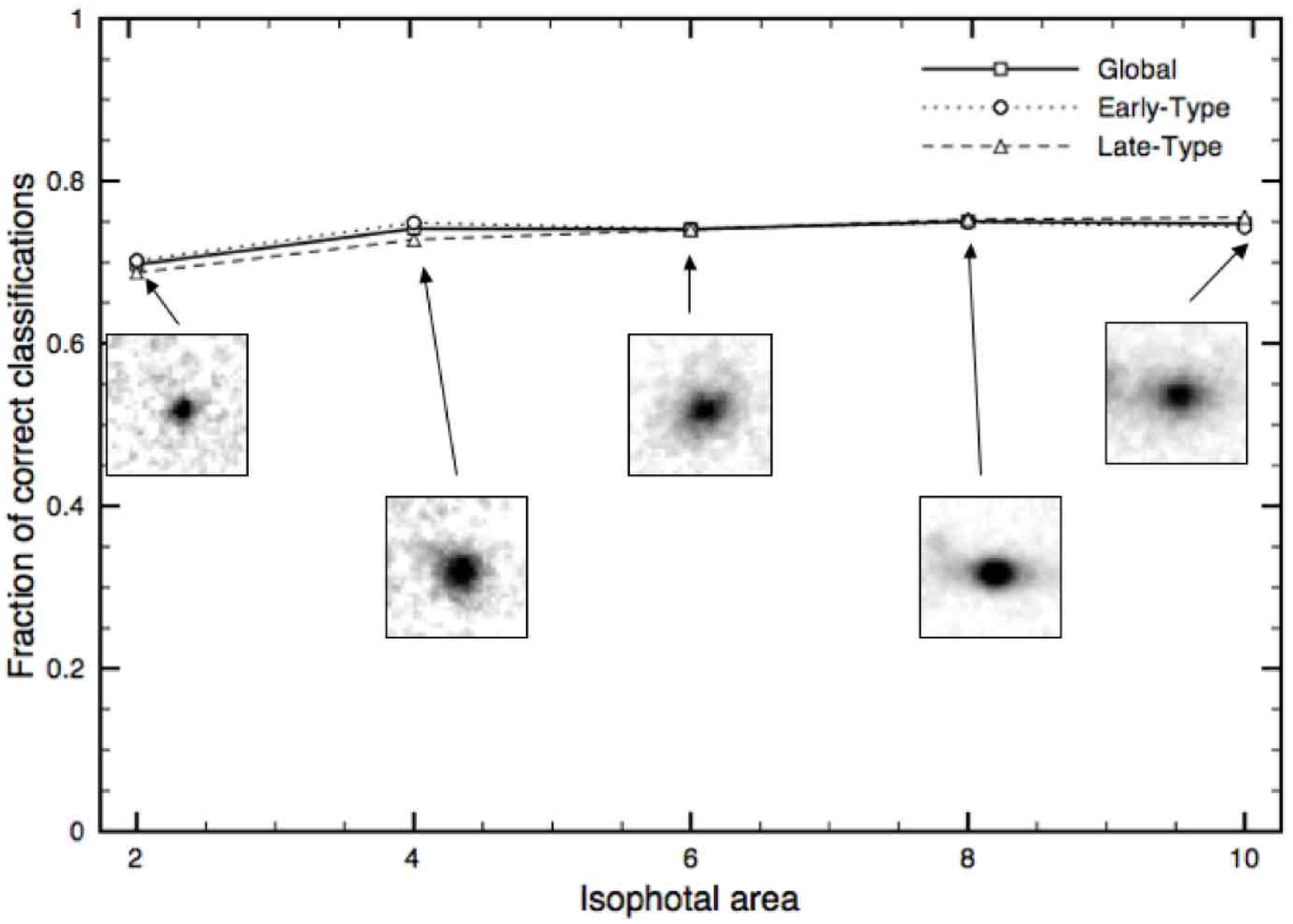} \\ 
	\mbox{\bf (c)} & \mbox{\bf (d)}\\
	\includegraphics[width=0.38\textwidth,height=0.3\textwidth]{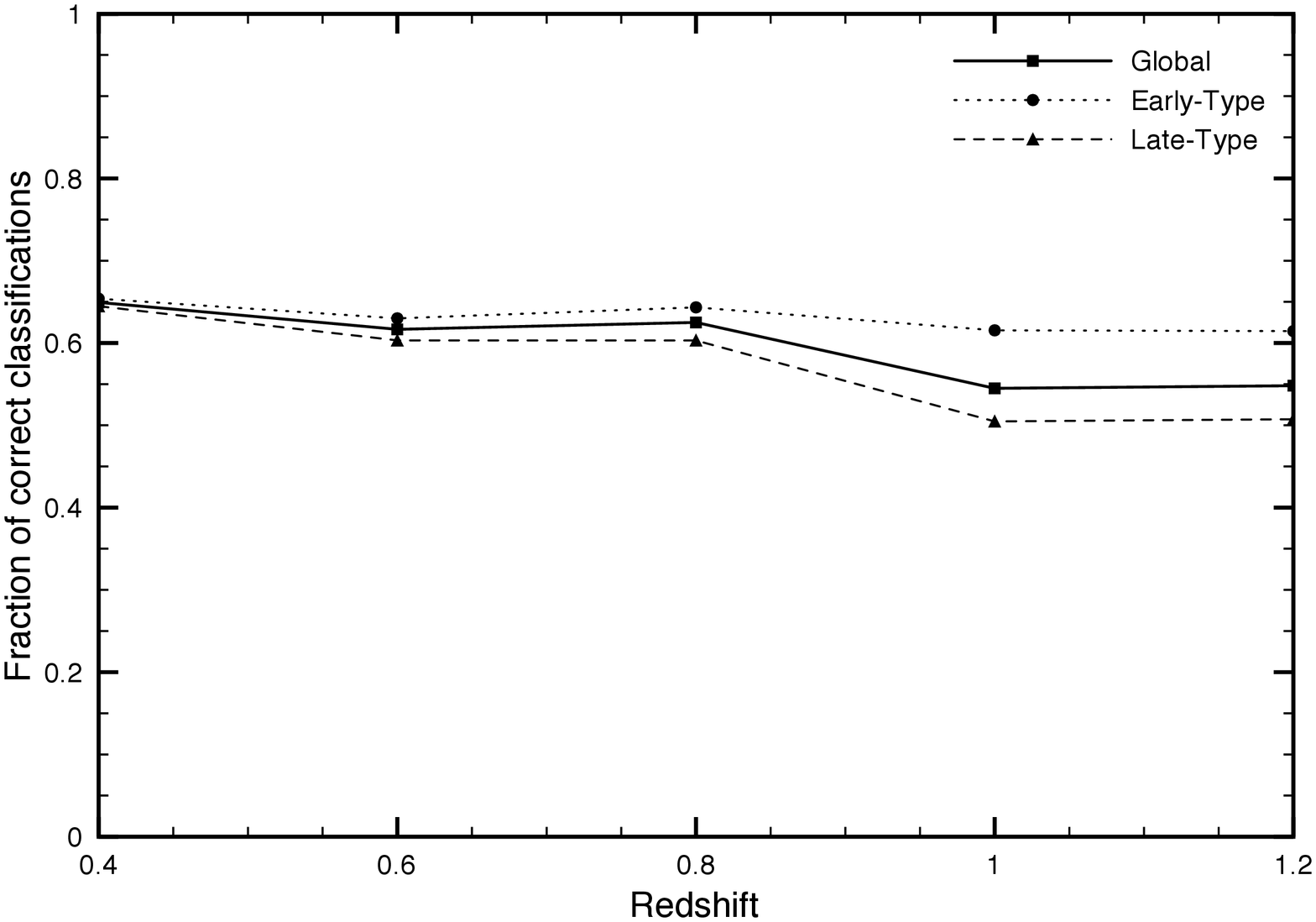} &
	\includegraphics[width=0.38\textwidth,height=0.3\textwidth]{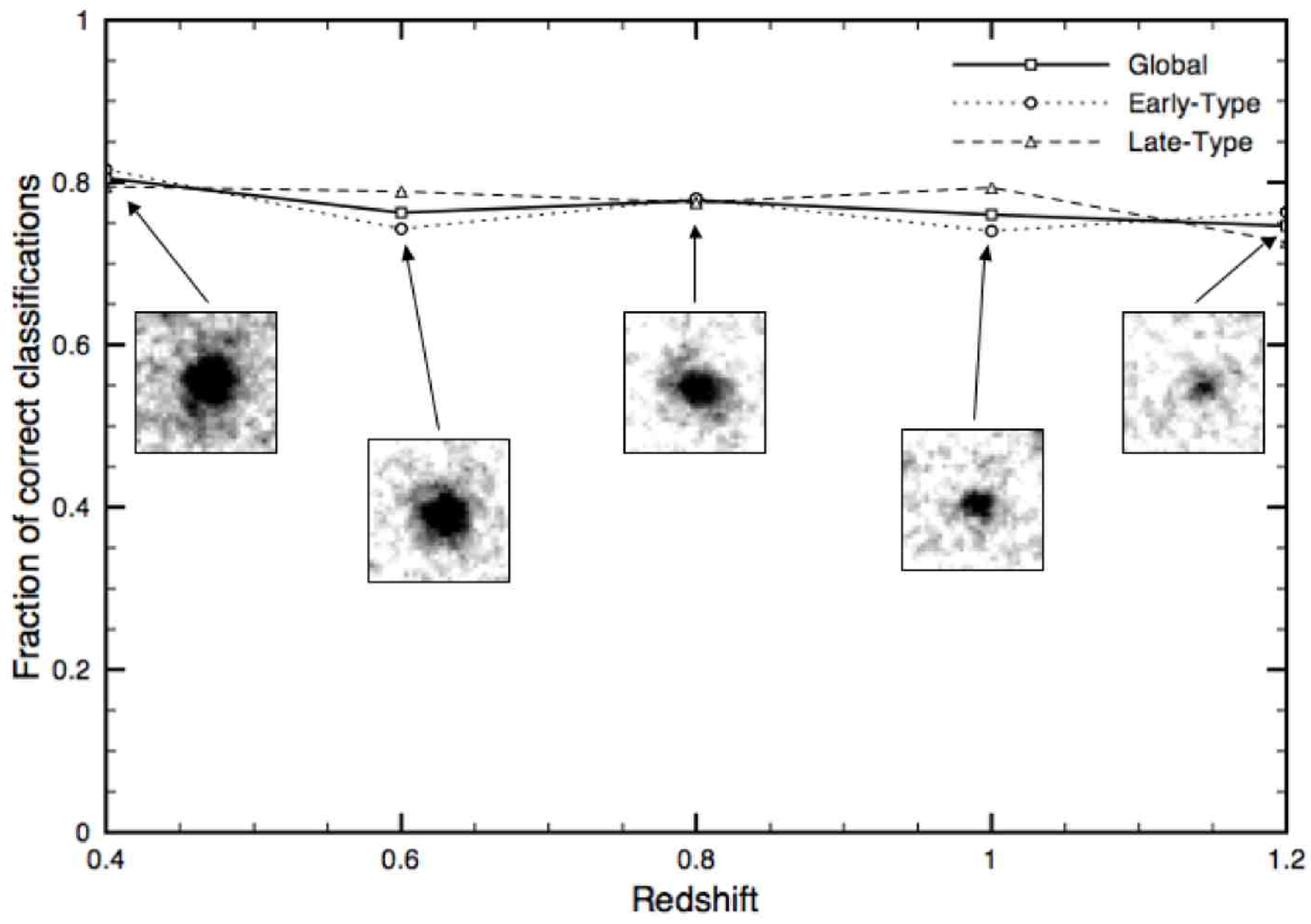} \\ 
	\mbox{\bf (e)} & \mbox{\bf (f)}\\
	

\end{array}$
\end{center}
\caption{Cumulative accuracy of classifications for a 2D machine (left column) and a 12D one (right column) as a function of magnitude (a and b), area (c and d) and redshift (e and f). Solid line shows the global accuracy, i.e. the number of galaxies correctly identified, dotted and dashed lines show respectively the fraction of early type and late type galaxies classified correctly. Stamps in the right column show a typical galaxy for every magnitude, area and redshift range.}
\label{fig:svm_comparison}
\end{figure*}
 For that purpose, we look at the accuracy of the classification as a function of 3 main properties of the galaxies: luminosity, distance and area (Fig.~\ref{fig:svm_comparison}) by progressively adding objects and measuring each time : a) the global accuracy, i.e. the fraction of galaxies that are classified correctly by the machine, and b) the accuracy per morphological type, i.e. the fraction of predicted early (late) type galaxies that are visually classified as early (late) type respectively ($N_{E\rightarrow E}$ and $N_{S\rightarrow S}$). 

Several conclusions can be extracted from this comparison:

\begin{itemize}

\item First, using more than two parameters simultaneously clearly increases the global accuracy of the classification in all the redshift, area or luminosity ranges. Indeed, the mean fraction of correct classifications in the 2 dimensional machine is around $\sim60\%$ and decreases to $\sim50\%$, which means that there is a high contamination in the C/A plane, whereas it rises up to more than $\sim80\%$ when using a 12 dimensional machine which is comparable of what is obtained in space observations \citep{Bri98,Men06}.

\item Second, the gain is even higher when looking at the $N_{E\rightarrow E}$ and $N_{S\rightarrow S}$ coefficients. For the C/A classification, there is indeed an asymmetric response of the machine: early type galaxies are better identified ($~65\%$) whereas the fraction of late type is significantly lower ($~50\%$), which means that an important fraction of late type galaxies are classified as early type. This will result, when doing the classification, in an important bias towards a too high fraction of elliptical galaxies. However, in  the 12 parameter classification, the accuracies are almost perfectly symmetric for the two morphological types.

\item Third, when looking at the evolution as a function of distance, size and luminosity, the 12 dimensions machine results in a more stable response in particular as a function of magnitude and redshift. 


\end{itemize}  



\subsubsection{How to fix the number of parameters?}

In section~\ref{sec:nd_2d} it is shown that the use of more than 2 dimensions to obtain morphology clearly increases the accuracy of the global classification. However, the questions that arise are: are all these parameters necessary? Might some parameters introduce a degeneracy and consequently reduce the machine's accuracy?

\begin{figure} 
 \centering 
  \resizebox{\hsize}{!}{\includegraphics{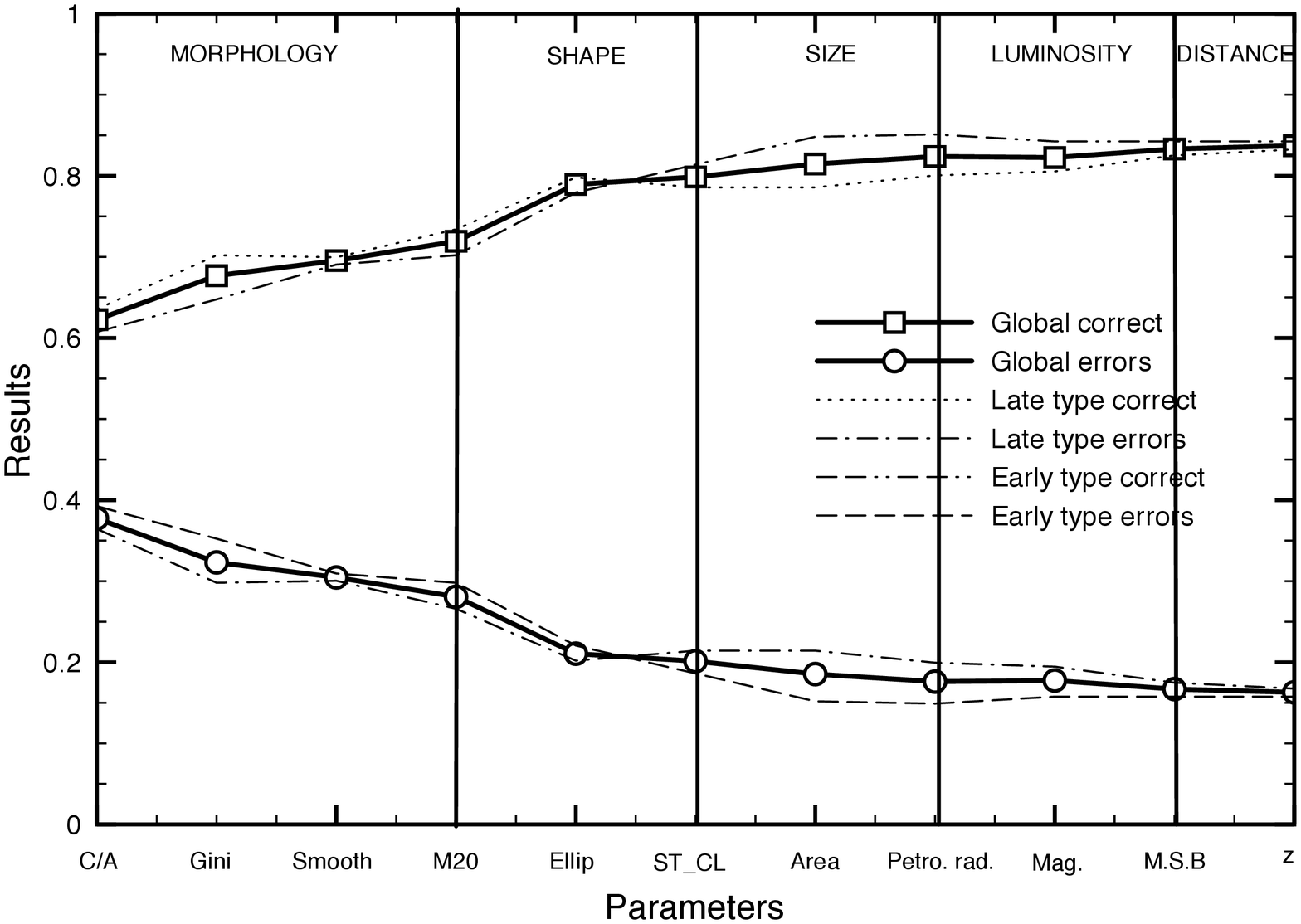}}
 \caption{Accuracy of the classification as a function of the number of parameters. The first point corresponds to a classical C/A classification and each new point adds a dimension. Parameters are classified in 5 classes: morphology, shape, size, luminosity, and distance.} 
 \label{fig:param_grow} 
 \end{figure}

To try to answer these questions we make a single test that consists in training several machines with an increasing number of parameters. We thus start with a classical 2 parameter machine (C and A) and we progressively add dimensions until we reach the 12 dimensions described in previous section. Results are plotted in Figure~\ref{fig:param_grow}. As above, we plotted the global accuracy and the one per morphological type. The dimensions are separated into 5 categories (morphology, shape, size, luminosity and distance). 

Two important points arise at first sight: first, not all the parameters bring the same amount of useful information. The morphology and the shape carry practically the necessary amount of information to reach $80\%$ accuracy. Second, the accuracy is a monotonic function of the number of parameters: adding a parameter can result in almost an unchanged accuracy (for instance, the magnitude) but never reduces it. This is particularly important, since it means that including more parameters than necessary does not result in a degeneracy. In addition, adding does not result in a significant increase of the computing time.

\subsubsection{Influence of the training set}

The method we adopted here for building the training set aims at
reproducing the observing conditions and physical properties of the
sample in order to reduce errors due to the difference between the
training and the science samples. The machine is thus trained to solve
a specific problem and should be trained differently for every science sample.
We now measure the importance of this effect by simulating the same
sample as if it was observed by the adaptive optics system NACO installed on the VLT. We use NACO data that have been observed in the $K_s$ band with 2
to 3 hours exposure time for each pointing \citep{Huer06}. The
total area covered by these data reaches 7 $arcmin^2$ and the mean
resolution is $0.1^{"}$. We therefore repeated the same procedure but
dropped the simulated catalogue in a real NACO background. We then
trained the machine with this sample and try to classify the WIRCam
simulated sample with the trained machine. 

Results are shown in table~\ref{tbl:models_comp}: the global accuracy of the classification falls to $62\%$, i.e. $40\%$ of contaminations when using the NACO model to classify WIRCam galaxies. In particular, there is a systematic drift from late to early type galaxies. The training set must thus be carefully built to take into account all the observing conditions.

\begin{table}[h!t!]
\begin{center}
\begin{tabular}{c|c|c}
\hline\hline\noalign{\smallskip}
& WIRCam Model & NACO Model \\
\noalign{\smallskip}\hline\noalign{\smallskip}
Global & $0.83$ & $0.62$ \\
$N_{E\rightarrow E}$ & 0.81 & 0.96 \\
$N_{S\rightarrow S}$ & 0.84 & 0.24 \\
\noalign{\smallskip}\hline
\end{tabular}
\caption{Accuracy of the classification when using a machine trained with a sample with different properties than the science sample - see text for details.}
\label{tbl:models_comp}
\end{center}
\end{table}

\section{Summary and conclusions}

We have presented a new method to perform morphological classification of cosmological samples based on support vector machines. It can be seen as a generalization of the classical non-parametrical C/A classification method but with an unlimited number of dimensions and non linear boundaries between the decision regions. The method is specially adapted to be used on large cosmological surveys since it is fully automated and errors are estimated objectively allowing an easy comparison between surveys with different properties. Furthermore, since the calibration sample is built from a nearby sample visually classified adapted to reproduce the physical and instrumental properties, the method can be even employed on seeing-limited data. Selecting the calibration sample in the high redshift sample's rest-frame turns the results robust towards wavelength dependent effects and makes it easier to interpret them in terms of evolution. \\

As a test, we use our method to classify a near-infrared seeing-limited sample observed with WIRCam at CFHT with a training set of $\sim 1500$ objects from the SDSS. We show that increasing the number of parameters in the analysis reduces errors by more than a factor 2; leading to a mean accuracy of $\sim 80\%$ of correct classification up to the sample completeness limit ($K_{AB}\sim22$). The accuracy is furthermore a monotonic function of the number of parameters.

The presented method is intended as a framework for future studies. In
particular, it can be used to 
look for luminosity and color evolution as a function of the
morphology. However this method is far more general and can be
appplied on many other samples of galaxies observed with ground-based
data with or without AO correction. Several applications will be
intended in order to study the effects of local environment and galaxy
density on the morphological evolution of galaxies both in the field
and in rich clusters of galaxies. The library is available for download at \url{http://www.lesia.obspm.fr/~huertas/galsvm.html}




\bibliographystyle{aa}
\bibliography{biblio.bib}

\end{document}